\documentstyle[epsfig,longtable]{aipproc}

\newcommand{\nn}{\nonumber}

\def\Dsl{\hbox{/\kern-.6000em D}} 

\def\dsl{\,\raise.15ex\hbox{/}\mkern-13.5mu D}
\def\bsigma{\mbox{\boldmath $\sigma$}}

\def\psip#1{\psi_{\mathbf{#1}}}
\def\chip#1{\chi_{\mathbf{#1}}}
\def\bsigma{\mbox{\boldmath $\sigma$}}
\def\OMIT#1{}
\def\mklrg#1{\mbox{\normalsize $#1$}}

\begin{document}

\title{Bound States in NRQCD/NRQED and the Renormalization
Group\footnote{\tighten Talk presented at the {\it 7th Conference on the
Intersections of Particle and Nuclear Physics}, Quebec City, Canada, 
May 22-28, 2000. UCSD/PTH 00-19}}  
\author{Iain W.\ Stewart}
\address{\tighten Department of Physics, University of California at San
Diego,\\[2pt] 9500 Gilman Drive, La Jolla, CA 92093 }

\maketitle

{\tighten
\begin{abstract}

The application of renormalization group techniques to bound states in
non-relativistic QED and QCD is discussed. For QED bound states like Hydrogen
and positronium, the renormalization group allows large logarithms of the
velocity, $\ln v$ (or equivalently $\ln\alpha$'s), to be predicted in a
universal and simple way. The series of $(\alpha \ln\alpha)$'s are shown to
terminate after a few terms. For QCD one can systematically sum infinite series
of the form $[\alpha_s \ln \alpha_s]^k$, and answer definitively the question
``$\alpha_s$ at what scale?''.

\end{abstract}
}
\vspace{-0.1in}

For Coulombic bound states of two fermions, the relevant scales include the mass
of the fermions, $m$, their momentum, $p\sim mv$, and their energy, $E\sim
mv^2$.  Since $v\ll 1$ it is useful to calculate properties of these bound
states in a double expansion in $v$ and $\alpha$, with terms $\alpha/v\sim 1$
kept to all orders.  However, the expansion may still involve large logarithms
of $v$ or $\alpha$, which appear through factors of $\ln(p/m)$, $\ln(E/p)$, and
$\ln(E/m)$. In this talk I discuss how the renormalization group can be used to
systematically predict and sum powers of $\alpha\ln
v$~\cite{LMR,amis,amis2,amis3,amis4,mss1}.

At a given order in QED, terms involving $\ln \alpha$ typically give the largest
contributions, and the precision of experiments make their prediction quite
important~\cite{exp}. In Ref.~\cite{amis4} it was shown for the first time that
$\ln\alpha$'s in the Lamb shift, hyperfine splittings, and annihilation decay
widths can be predicted with the renormalization group.  This is in contrast to
the usual method of computing these logarithms by evaluating matrix elements at
the scale $m$.  The calculations are simple enough that we can simultaneously
treat Hydrogen, muonium ($\mu^+e^-$), and positronium ($e^+e^-)$.

For the $t\bar t$ system near threshold, the relevant scales are $m_t\sim
175\,{\rm GeV}$, $m_tv\sim 30\,{\rm GeV}$, and $m_tv^2\sim 5\,{\rm GeV}$, which
are all $\gg \Lambda_{\rm QCD}$ and can be treated perturbatively.  In QCD there
is a strong dependence of the coupling on the scale; $\alpha_s(m_t)$ is much
different from $\alpha_s(m_tv^2)$.  The renormalization group allows us to
handle this complication, or equivalently it allows us to systematically sum
terms $(\alpha_s \ln v)^k$.  Predictions for the running coefficients of the
$t\bar t$ potentials and the $t\bar t$ production current will be discussed
below~\cite{amis,amis3}.

\begin{figure}[!t]
  \centerline{\hspace{2.5cm} \epsfxsize=6.3truecm 
  \epsfbox{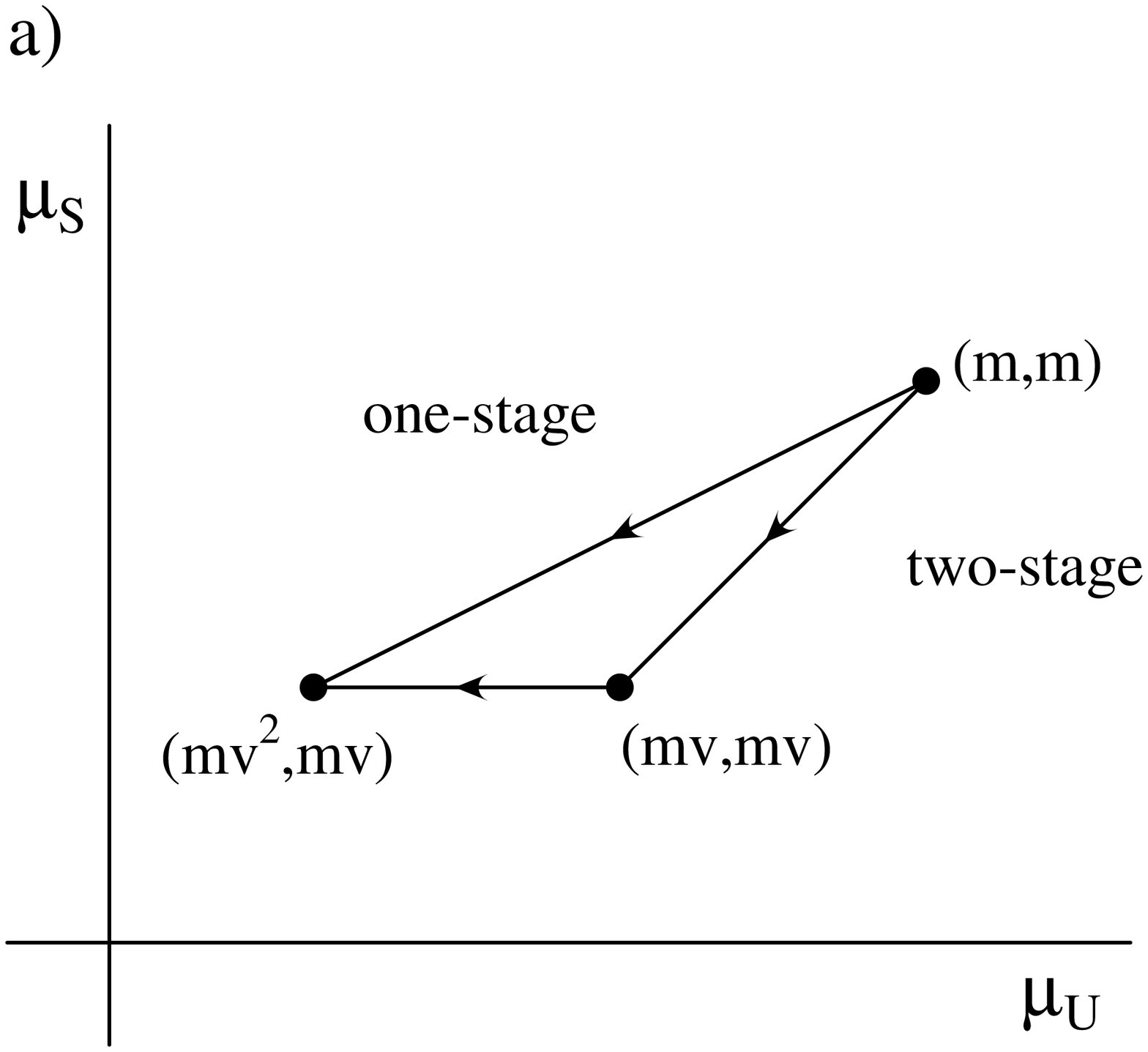} \ 
  \epsfxsize=10.0truecm \raise1.cm \hbox{
  \epsfbox{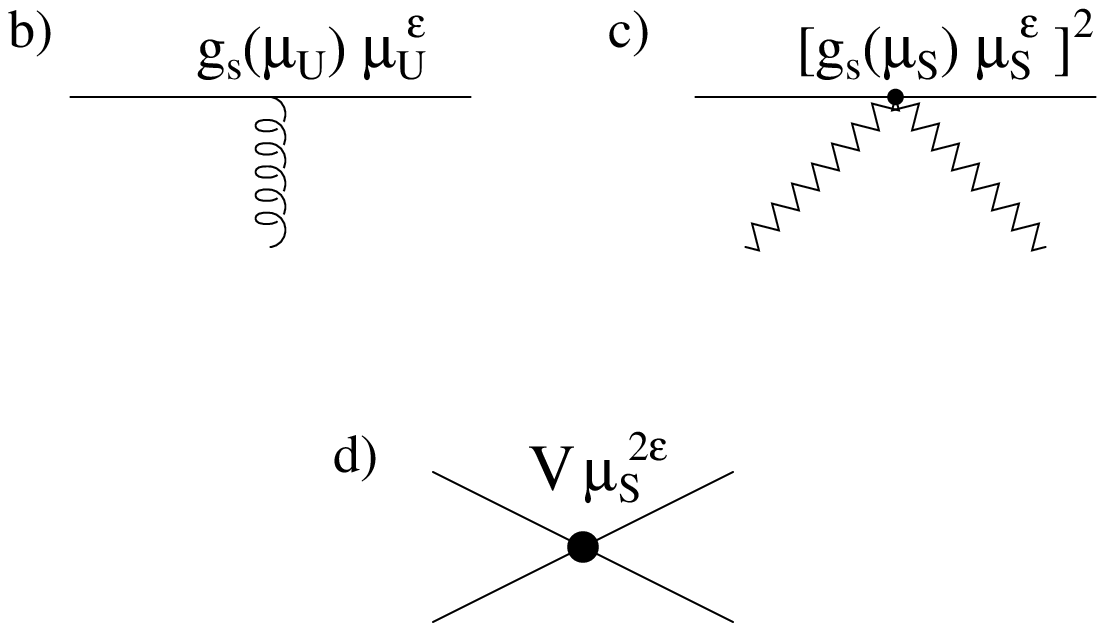}} } \vspace{0.3cm} 
\caption{a) Paths in the $(\mu_U,\mu_S)$ plane for one-stage and two-stage 
running. b),c),d) Examples of the $\mu_U$ and $\mu_S$ dependence of the 
Feynman rules.}
\label{fig_path}
\end{figure}

Effective theories for non-relativistic QED and QCD (NRQED and NRQCD) allow the
double expansion in $v$ and $\alpha$ to be performed in a simple
way\cite{Caswell}.  However, application of the renormalization group in
these theories is complicated by the presence of two low energy scales, $p$ and
$E$, which are coupled by the equations of motion, $E=p^2/(2m)$. If one attempts
to lower the cutoff on the energy to $E\lesssim \Lambda$, then one can still
excite larger momenta $p\lesssim \sqrt{m\Lambda}$.  Using dimensional
regularization one can deal with this coupling of scales by using a velocity
renormalization group~\cite{LMR}, which has a subtraction velocity $\nu$ rather
than the usual subtraction momentum $\mu$. Running in one-stage from $\nu=1$ to
$\nu=v$ simultaneously lowers the subtraction point for momenta, $\mu_S\equiv
m\nu$, to the scale $mv$, and the subtraction point for energy, $\mu_U\equiv
m\nu^2$, to $mv^2$.  In Fig.~\ref{fig_path}a this one-stage approach is
contrasted with the alternative two-stage approach where one first runs from
$\mu=m$ to $mv$ and then runs from $\mu=mv$ to $mv^2$.

In Ref.~\cite{mss1} it was shown that for QED bound states, the most obvious
method of two-stage running fails to reproduce terms involving $(\ln\alpha)^k$
with $k\ge 2$.  This occurs because in the two-stage method the coupling between
the energy and momentum is ignored.

To calculate observables at the hard scale $m$, the way the effective theory is
formulated does not matter too much as long as it has a consistent power
counting in $v$.  Below $m$, only on-shell degrees of freedom are kept in the
effective theory, i.e. degrees of freedom which fluctuate near their mass
shell. For example, a potential gluon exchanged between two quarks has energy
$\sim mv^2$ but momentum $\sim mv$, and is therefore far offshell.  Instead of a
potential gluon the effective theory has a four quark operator
\begin{eqnarray} \label{Lp}
 {\cal L}_p= - \sum_{\mathbf p,p'} V\left({\bf p},{\bf p^\prime}\right)\  
 \mu_S^{2\epsilon}\ {\psip {p^\prime}}^\dagger\:
  {\psip p}\: {\chip {-p^\prime}}^\dagger\:  {\chip {-p}}{} \,,
\end{eqnarray}
where $\psi_{\bf p}$ ($\chi_{\bf p}$) destroys a quark (antiquark) with momentum
${\bf p}$ and spin and color labels are suppressed. $V({\bf p,p^\prime})$
is the potential
\begin{eqnarray}
 V({\bf p,p^\prime}) &=& { U_c \over {\bf k}^2} + { U_k \over |{\bf k}| } + U_2 
  + U_s {\bf S}^2 \,+ { U_r\: ({\bf p^2 +p^{\prime 2}}) \over 2\,{\bf k}^2} 
  +  U_t\:  \bigg( \bsigma_1\cdot \bsigma_2 - {3\, {\bf k} \cdot \bsigma_1 
  {\bf k}\cdot \bsigma_2 \over {\bf k}^2 } \bigg)  \nn\\ 
 && -  {i {\bf U_\Lambda} \cdot ( {\bf p\,^\prime \times p} ) 
  \over {\bf k}^2 } + { {{U}_3}\,|{\bf k}|  } 
  + { {{U}_{3s}}\,{\bf S^2} |{\bf k}|   } +
  { {{U}_{rk}}\,({\bf p^2}+{\bf p'^2}) \over  2|{\bf k}| } 
  + \ldots \,,
\end{eqnarray}
where the $U_i(\nu)$'s are running coefficients. In QCD the color singlet and
octet coefficients have different values.  We have absorbed in the $U_i$'s the
dependence on the fermion masses; only the momentum dependence is important for
the power counting.  For the $QQ$ and $Q\bar Q$ potentials in an arbitrary Lie
gauge group, the matching coefficients at one-loop to order $v^2$ can be found in
Ref.~\cite{amis2}.

The effective Lagrangian has quarks with $(E,p)\sim (mv^2,mv)$ interacting with
soft gluons which have $(E,p)\sim (mv,mv)$, and ultrasoft gluons which have
$(E,p)\sim (mv^2,mv^2)$ (see Refs.~\cite{LMR,amis,amis2,amis3}).  To implement
the velocity renormalization group we renormalize the Lagrangian and compute
anomalous dimensions. This procedure is fairly simple since it can be done
treating the Coulomb potential perturbatively.  In Figs.~\ref{fig_path}b,c,d
the $\mu$'s which appear in some typical interactions are shown.
Fig.~\ref{fig_path}b shows a quark interacting with a single ultrasoft gluon.
Physically, the fact that the ultrasoft mode involves the coupling $g(\mu_U)$
makes sense; due to the multipole expansion the scale $mv^2$ is the only scale
it sees.  Fig.~\ref{fig_path}c shows a soft gluon scattering off a quark, and
Fig.~\ref{fig_path}d shows an insertion of the potential. For these
interactions the parameter $\mu_S\sim mv$ appears.

\begin{table}[t!]  
\caption{QED $\ln\alpha$'s which follow from the leading order (LO) and
next-to-leading order (NLO) anomalous dimensions in Ref.~\protect\cite{amis4}.
$\mu^+e^-$ predictions include $1/m_\mu$ dependence. (h.f.s is hyperfine
splitting and $\Delta\Gamma/\Gamma$ is the $e^+e^-$ decay width correction.)}
\begin{tabular}{l|cclcl}
\noalign{\vspace{-8pt}}
 $\alpha^8\ln^3\alpha$ & Lamb shift && $H$ & & 
  agrees with \cite{karshenboim,goidenko},\\ &&&&&  disagrees with 
  \cite{ms,yerokhin}\tablenote{There is a growing consensus that the value 
  in Ref.~\cite{karshenboim} is correct~\cite{Sapirstein}.} \\[2pt] 
  & && $\mu^+e^-$, $e^+e^-$  && new predictions \\
  & (no h.f.s.)  \\[2pt]
 $\alpha^4\ln^3\alpha$ & (no $\Delta\Gamma/\Gamma$) & & \\[5pt] \hline
 $\alpha^7\ln^2\alpha$ & Lamb shift && $H$, $\mu^+e^-$, $e^+e^-$ && needs
 running of $V^{(1)}$ \\
  & h.f.s. && $H$, $\mu^+e^-$, $e^+e^-$ && agree with 
  \cite{thesis,karshenboim,melnikov} \\
 $\alpha^3\ln^2\alpha$ & $\Delta\Gamma/\Gamma$ && $e^+e^-$ ortho and para && 
  agree with \cite{karshenboim} \\[5pt] \hline
 $\alpha^6\ln\alpha$ & Lamb shift, h.f.s. &&  $H$, $\mu^+e^-$, $e^+e^-$ &&
  needs $\rho_s$, $V^{(1)}(1)$ \\
 $\alpha^2\ln\alpha$ & $\Delta\Gamma/\Gamma$ && $e^+e^-$ ortho and para && 
  agree with \cite{caswell2} \\[5pt] \hline\hline
 $\alpha^5\ln\alpha$ & Lamb shift && $H$, $\mu^+e^-$, $e^+e^-$ && agree \\
   & (no h.f.s.) & \\
 $\alpha\ln\alpha$  & (no $\Delta\Gamma/\Gamma$) & 
\end{tabular} \label{lnas}
\end{table} 

Below the electron mass the electromagnetic coupling in NRQED does not run, but
coefficients in the potential do. For fermions with mass and charge $(m_1,-e)$ 
and $(m_2,Ze)$ we find the anomalous dimensions: 
\begin{eqnarray}\label{NLO}
   \left .\nu {d U_2 \over d \nu}\right|_{\rm LO} &=& {2\alpha \over 3 \pi}
  \left({1\over m_1}+ {Z \over  m_2}\right)^2 U_c + {14 Z^2 \alpha^2 \over 
  3 m_1 m_2} \,, \nn \\
   \left. \nu {d U_{2+s} \over d\nu}  \right|_{\rm NLO} \!\!\!&=&  
  \rho_{c22}\, U_c\left(U_{2+s}^2 +2 U_{2+s} U_r + \frac34 U_r^2  
  -9 U_t^2 {\bf S^2} \right)+ \rho_{ccc}\, U_c^3 \nn\\*
 &&\quad + \rho_{cc2}\, U_c^2  \left( U_{2+s}+ U_r \right)
    +\rho_{ck}\, U_c U_k +\rho_{k2}\, U_k \left(U_{2+s}  +
   U_r/2\right) \nn\\*
 && \quad + \rho_{c3}\, U_c \left(U_3+U_{3s} {\bf S^2}+\frac{U_{rk}}{2} \right)
   + \rho_s {Z^3 \alpha^3 \over m_1 m_2},
\end{eqnarray} 
where $U_{2+s}=U_2+U_s {\bf S^2}$ and the $\rho_i$'s are mass dependent
numbers\cite{amis4}. Solving these equations gives the results summarized in
Table~\ref{lnas}.\footnote{Note that the $\alpha^7\ln^2\alpha$ Lamb shift
requires the LL running of $V^{(1)}$ since this potential mixes into $U_2$, and
the $\alpha^6\ln\alpha$ Lamb shift requires the NLL running of $U_k$. These will
be discussed in a future publication.}  Taking the matrix element of the leading
log (LL) value of $U_2(\nu)$ gives the $\alpha^5\ln\alpha$ Lamb shifts for
Hydrogen, muonium, and positronium. Furthermore, the LO anomalous dimension is
independent of $\nu$ so there are no higher terms, $\alpha^{k+4}\ln^k\alpha$ for
$k\ge 2$.  At next-to-leading log (NLL) order the most logarithms are generated
by the $\rho_{c22} U_c U_2(\nu)^2$ term which gives the $\alpha^8 \ln^3\alpha$
Lamb shifts. Thus, there are no terms $\alpha^{k+5}\ln^k\alpha$ for $k\ge
4$. Hyperfine splittings are generated by the ${\bf S^2}$ terms in
Eq.~(\ref{NLO}), and the ortho and para-positronium widths are generated by
imaginary terms which enter through the matching condition for $U_{2+s}(1)$.

\begin{table}[t!]  
\caption{LO and LL values of coefficients of the $t\bar t$ potential in QCD.
Here ${\bf U_\Lambda}={\bf S} U_\Lambda$ and the values are in units of the top
mass $m_t$.}
\begin{tabular}{l|cccccc}
\noalign{\vspace{-8pt}}
 & \mklrg{U_k^{(s)} m_t} & \mklrg{U_r^{(s)}m_t^2} & \mklrg{U_2^{(s)}m_t^2} 
  & \mklrg{U_s^{(s)}m_t^2} & \mklrg{U_\Lambda^{(s)}m_t^2}
  & \mklrg{U_t^{(s)}m_t^2} \\[3pt] \hline
 $\ \mklrg{\nu=1}\ $ & \mklrg{-0.36} & \mklrg{-1.81} & \mklrg{0} & 
  \mklrg{0.60} & \mklrg{0.15} & \mklrg{2.71} \\
  $\ \mklrg{\nu=v}\ $ & \mklrg{-0.03} & \mklrg{-1.49} & \mklrg{0.63}  & 
    \mklrg{0.53} & \mklrg{0.16} & \mklrg{3.11} 
\end{tabular} \label{Vs}
\end{table} 
NRQCD is better at generating logarithms than NRQED since the running of
$\alpha_s$ causes all potential coefficients to run.  For $t\bar t$ the change
in the color singlet couplings from $\nu=1$ to $\nu=v=0.15$ are shown in
Table~\ref{Vs}.\footnote{The renormalization group improved static potential is 
considered in Ref.~\cite{static}.}  The largest changes occur in the spin 
independent couplings
$U_k^{(s)}(\nu)$, $U_r^{(s)}(\nu)$, and $U_2^{(s)}(\nu)$. It is these couplings
which depend on $\alpha_s(m_t\nu^2)$ since their anomalous dimensions have
contributions from ultrasoft diagrams.  In Fig.~\ref{Vs}a we plot the two-loop
running of $U_k^{(s)}(\nu)$ whose value changes by an order of magnitude between
$\nu=1$ and $\nu=v$.
\begin{figure}[!t]
  \centerline{\epsfxsize=7.1truecm 
  \epsfbox{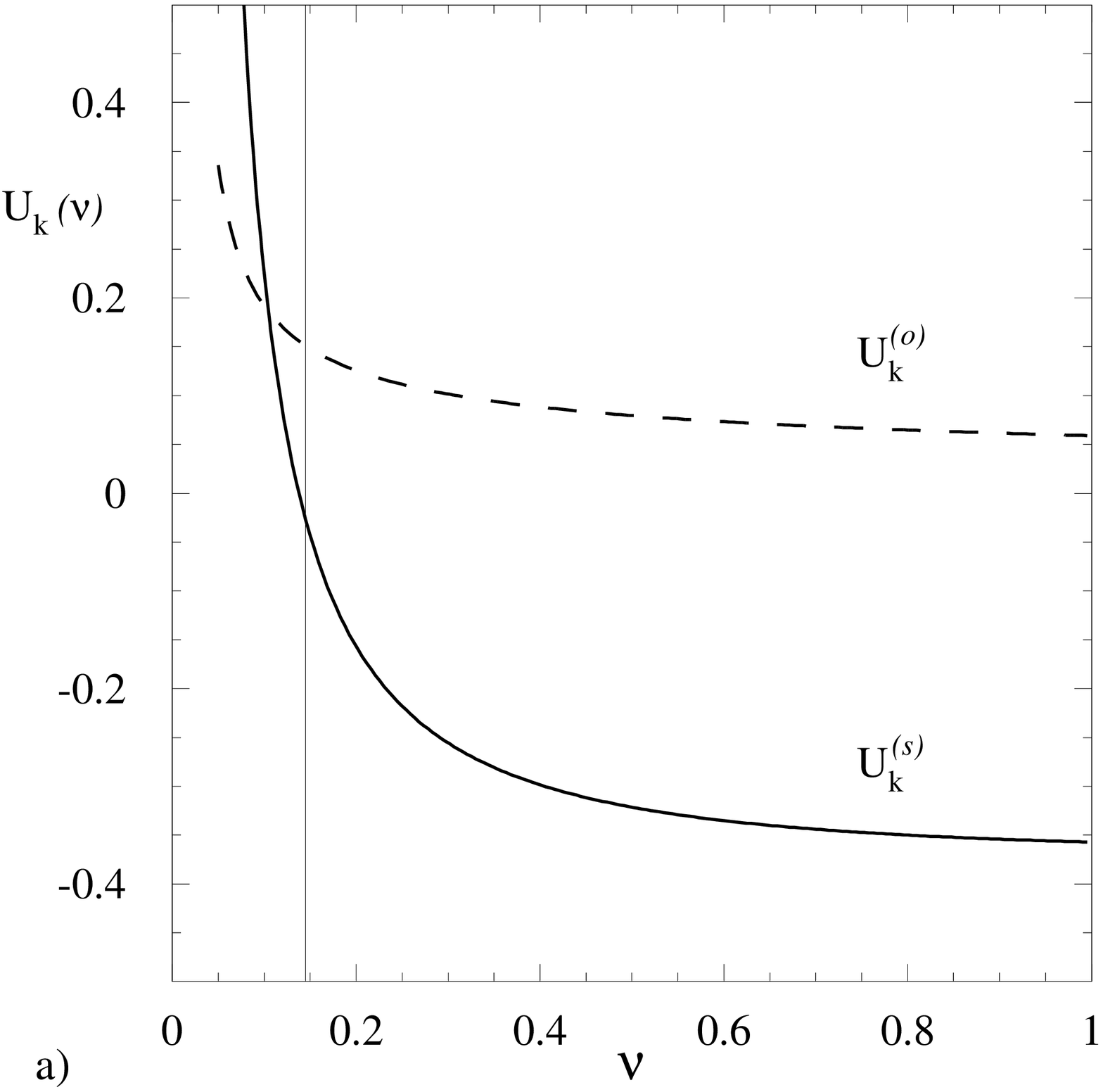}\ \ \ \
  \epsfxsize=7.1truecm 
  \epsfbox{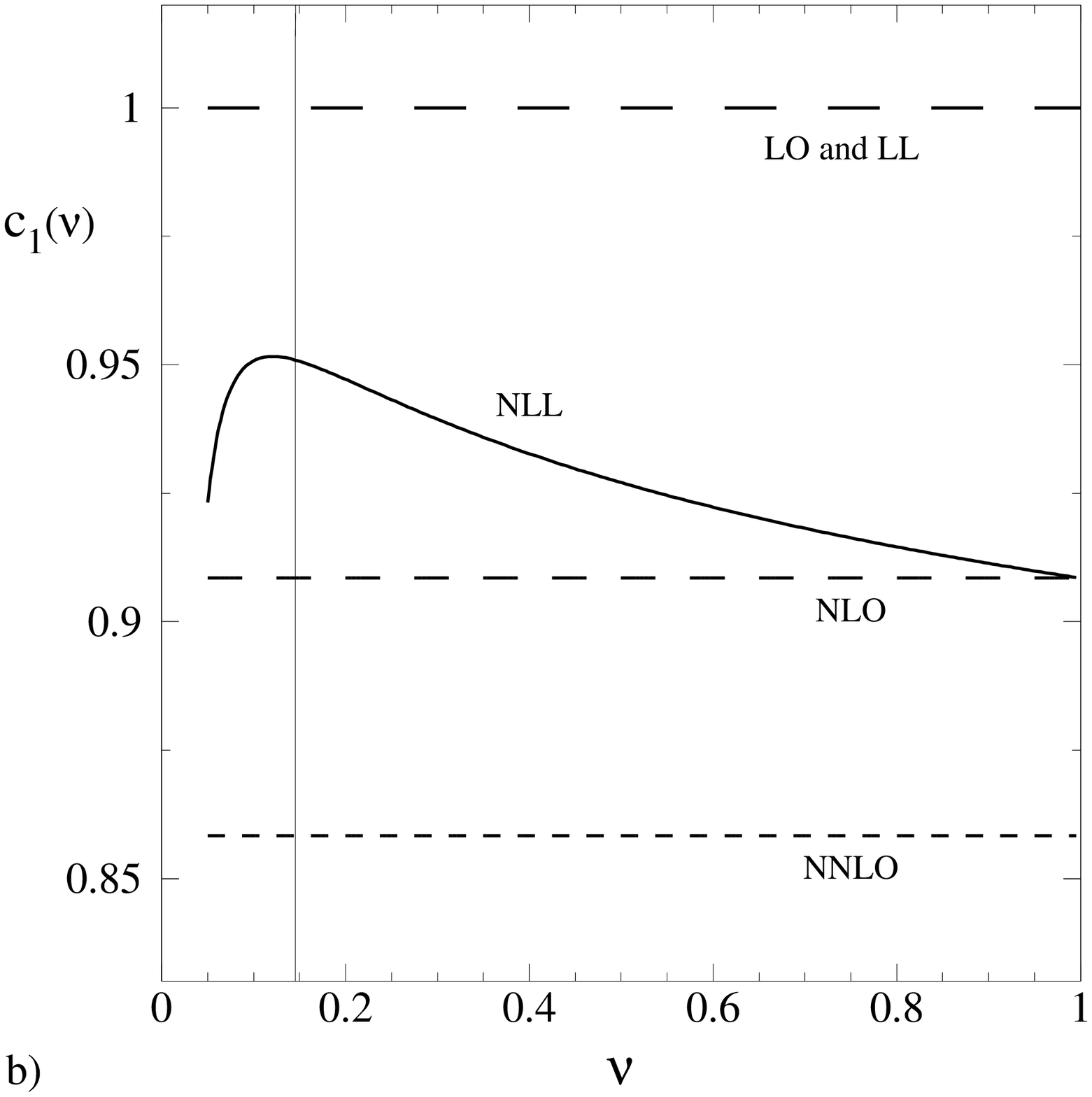} } \vspace{0.3cm} 
\caption{For $t\bar t$ production near threshold the running of the color
singlet and octet $1/|{\bf k}|$ potentials are shown in a), and the NLL value of
the production current is given in b) (solid line) \protect\cite{amis3}. In b)
the large, medium, and small dashes are the LO, NLO, and
NNLO\protect\cite{prodcurrent} matching results. In a) and b), $\nu=1$ is the
scale $\mu=m_t$ and the solid vertical line is the Coulombic velocity.}
\label{fig_vk}
\end{figure}
The full theory $t\bar t$ production current gets matched onto a current in the
effective theory $\bar t\gamma^i t = c_1 \psi^\dagger_{\bf p} \sigma^i
\chi^*_{\bf -p}+\ldots$.  At NLL order the running potentials mix into the
running of the production current coefficient $c_1(\nu)$\cite{LMR}.
Fig.~\ref{Vs}b plots the running of $c_1(\nu)$ at NLL from Ref.~\cite{amis3}.
Summing the logarithms improves the convergence by reducing the size of the NLO
matching coefficient by a factor of $2$.  It would be interesting to see if a
similar improvement in the convergence takes place for the rather large NNLO
matching correction found in Ref.~\cite{prodcurrent}.

This work was supported in part by the Department of Energy under grant
DOE-FG03-97ER40546, by the National Science Foundation under award NYI
PHY-9457911, and by NSERC of Canada.

{\tighten

} 

\end{document}